# Spin-Dependent Quantized Magnetic Flux Through The Electronic Orbits of Dirac Hydrogen Atom

M. Saglam[a], B. Boyacioglu[b], Z. Saglam[c], O. Yilmaz[d] and K. K. Wan[e]

[a]*Ankara University, Department of Phyics, Tandogan, Ankara, Turkey*
[b]*Ankara University, Dikimevi Vocational School of Health, Kecioren, Ankara, Turkey*
[c]*DLH Research Lab. Ministry of Transportation, Macunkoy, Ankara, Turkey*
[d]*Canakkale Onsekiz Mart University, Department of Physics, Canakkale, Turkey*
[e]*St. Andrews University, School of Physics and Astronomy, Scotland, UK*

**Abstract**

We investigate the quantized magnetic flux through the electronic orbits of Dirac hydrogen atom in the absence of an external magnetic field. The sources of the magnetic fields are taken to be that of proton's magnetic moment $\mu_p$ and electron's magnetic moment $\mu_e$ (or $\mu_j$) which has two components namely the orbital part $\mu_l$ and the spinning part $\mu_s$. We show that the quantized magnetic fluxes through the electronic orbits corresponding to the $(n, l=n-1, m_j)$ eigenstates of Dirac hydrogen atom take the forms: $\Phi(n,l,m_j) = [n-l-m_j]\Phi_0 = [n-l-m_l-m_s]\Phi_0$ where $\Phi_0 = \frac{h}{|e|}$ is the flux quanta. The application of the present result to the selection rules for the optical transitions of hydrogen atom gives access to the spin flip-floppings. The present result is believed to serve a significant help for understanding the recent observations of spin relaxation in excitonic transitions (such as 1s→2p or 2p→3d) in nanostructures.

Corresponding Author : Mesude Saglam
Adress : Ankara University, Department of Physics, 06100-Tandogan, Ankara, Turkey.
e-mail : saglam@science.ankara.edu.tr
Tel    : + 90 312 2126720, Ext.: 1222
Fax    : + 90 312 2232395

# 1. Introduction

Magnetic flux quantization was first recognized by London [1] and Onsager [2] who argued that in the interior of the superconductor the current density vanishes and the requirement that the electron wave function be single valued results the enclosed flux through a superconducting ring is quantized in units of $\frac{h}{|e|}$. Deaver and Fairbank [3] observed experimentally quantized values of magnetic flux through a superconducting ring. Onsager, L. and Lifshitz, I. M. used the Bohr-Sommerfeld phase integral method to show that the orbit of the electron in a uniform magnetic field is quantized in such a way that [4] the flux through it is $\Phi_n = (n+\frac{1}{2})\Phi_0$ (n = 0,1,2,3,...) where $\Phi_0 = \frac{h}{|e|}$. Recently Saglam and Boyacioglu [5] argued that there will be a spin contribution to the magnetic flux $\Phi_s = (\pm\frac{1}{2})\Phi_0$ which is based on the magnetic top model [6-8]. As was argued in [5] the quantized magnetic flux can be obtained through the conversion of the area integral of the magnetic field to a time integral over the cyclotron period $T_c$ $(=\frac{2\pi}{\omega_c} = \frac{2\pi m_e}{eB})$. Therefore for calculating the quantized magnetic flux the important time interval was the cyclotron period of the magnetic field. The aim of the present study is to calculate the quantized magnetic fluxes through the electronic orbits of Dirac hydrogen atom in the absence of an external magnetic field. The sources of the magnetic fields are taken to be that of proton's magnetic moment $\mu_p$ and electron's magnetic moment $\mu_e$ (or $\mu_j$) which has two components, namely the orbital part $\mu_l$ and the spinning part $\mu_s$. We show that the quantized magnetic fluxes through the electronic orbits corresponding to the $(n,l=n-1,m_j)$ eigenstates take the forms: $\Phi(n,l,m_j) = [n-l-m_j]\Phi_0 = [n-l-m_l-m_s]\Phi_0$ where $\Phi_0 = \frac{h}{|e|}$ is the flux quanta. Further we ask the requirement that between which states the flux change ($\Delta\Phi$) is equal to $(0, \pm\Phi_0)$. The answer is that for those states that satisfy: ($\Delta m_j = 0, \pm 1$). We notice that this condition is exactly the same spin dependent selection rules as obtained recently by Saglam et. al. [9] for optical transitions in Dirac hydrogen atom using the well-known Golden rule [10]. We show that both results allow spin flip which is not possible according to the conventional selection rules. As a simple example we look at the 1s → 2p optical transitions and again showed that some transitions allow spin flip which is not possible according to the conventional selection rules. The present result can serve a significant help for understanding the recent observation of the spin relaxation in excitonic transitions in nanostructures [11-13]. The outline of the present study is as follows: In section 2 we give a brief summary of the Schrödinger hydrogen atom. Section 3 deals with the Dirac-Hydrogen atom. In section 4 we calculate the quantized magnetic fluxes through the electronic orbits of the hydrogen atom. In Section 5 we apply the present result to 1s → 2p optical transitions. Section 6 contains the conclusions.

**2. The Schrödinger Hydrogen Atom**

In the hydrogen atom the potantial energy $V(r) = -\frac{e^2}{r}$ is one of the simplest potantial in quantum mechanics that can be solved analytically. Although the problem is a two body problem the related wave equation becomes one particle equation after the center of mass motion is separated out. Because of the fact that proton is more massive than electron, we can also assume that the proton is at rest at the origin of the center of mass system. In this approximation the non-relativistic Schrödinger equation for electron reads [14-16]:

$$H_0 \Psi_{n,l,m} = \left[ \frac{p^2}{2m} - \frac{e^2}{r} \right] \Psi_{n,l,m} = E_{n,l,m} \Psi_{n,l,m} \qquad (1)$$

where the first term in square bracket is the kinetic energy operator and the second term is the potantial energy of the electron in the rest mass frame of the proton. In spherical coordinates the potential energy in Eq.(1) produces the following results:

$$|n,l,m> = \Psi_{n,l,m}(r,\theta,\phi) = R_{nl}(r)Y_l^m(\theta,\phi), \qquad (2)$$

namely, the wave function of the electron is the multiplication of a radial part $R_{n,l}(r)$ and an angular part $Y_l^m(\theta,\phi)$ which are called spherical harmonics and take the form:

$$Y_l^m(\theta,\phi) = (-1)^m \left[\frac{2l+1}{4\pi}\frac{(l-m)!}{(l+m)!}\right] P_l^m(\cos\theta)e^{im\phi} \qquad (3)$$

where $P_l^m(\cos\theta)$ are the Legendre functions. Orthonormality relation for $Y_l^m(\theta,\phi)$ functions is:

$$\int_0^\pi \int_0^{2\pi} [Y_{l'}^{m'}(\theta,\phi)]^* Y_l^m(\theta,\phi) \sin\theta d\theta d\phi = \int (Y_{l'}^{m'})^* Y_l^m d\Omega = \delta_{l'l}\delta_{m'm} \qquad (4)$$

The radial wave function, $R_{n,l}(r)$ satisfies the radial Schrödinger equation given by

$$\left(\frac{d^2}{dr^2} + \frac{2}{r}\frac{d}{dr}\right)R_{n,l}(r) + \frac{2m_e}{\hbar^2}\left(E_{n,l,m} + \frac{e^2}{r} - \frac{l(l+1)}{2m_e r^2}\hbar^2\right)R_{n,l}(r) = 0 \qquad (5)$$

The solutions of Eq.(5) are given in terms of Laguerre polynomials:

$$R_{nl}(r) = \left(\frac{2}{na_0}\right)^{3/2}\left\{\frac{(n-l-1)!}{2n[(n+l)!]^3}\right\}^{1/2}\rho^l e^{-\rho/2}L_{n+1}^{2l+1}(\rho) \qquad (6)$$

where $a_0 = \frac{\hbar^2}{m_e e^2}$ is the Bohr radius and $\rho = \frac{2r}{na_0}$. Here $m_e$ is the mass of the electron

The energy eigenvalues $E_{n,l,m}$ in Eq.(1) only depend on the principal quantum number $n$, but not depend on $l$ and $m$:

$$E_{n,l,m} = E_n \propto \frac{1}{n^2}. \qquad (7)$$

## 3. The Dirac Hydrogen Atom

Let us begin with the Dirac Hamiltonian of the hydrogen atom [9,10,14-16]

$$H_D = \alpha \cdot pc + \beta m_e c^2 + V(r) \tag{8}$$

where $V(r)$ is the Coulomb potential, $m$ is the mass of an electron, $c$ is the velocity of light and $\alpha$ and $\beta$ are the standard Dirac matrices in the Dirac representation

$$\alpha = \begin{pmatrix} 0 & \sigma \\ \sigma & 0 \end{pmatrix} \qquad \beta = \begin{pmatrix} 1 & 0 \\ 0 & -1 \end{pmatrix} \tag{9}$$

Here the 1's and 0's stand, respectively, for 2×2 unit and zero matrices and the $\sigma$ is the standard vector composed of the three Pauli matrices $\sigma = (\sigma_x, \sigma_y, \sigma_z)$. Since the Hamiltonian (is invariant under rotations, we look for simultaneous eigenfunctions of $H_D$, $|\mathbf{J}|^2$, and $J_z$, where

$$\mathbf{J} = \mathbf{L} + \mathbf{S} \quad ; \quad S \equiv \frac{1}{2}\Sigma = \frac{\hbar}{2}\begin{pmatrix} \sigma & 0 \\ 0 & \sigma \end{pmatrix} \tag{10a}$$

and we write

$$J_z = L_z + S_z \quad \text{or} \quad m_j = m_l + m_s \equiv m + m_s. \tag{10b}$$

We note that the spin operator is diagonal in terms of 2×2 Pauli spin matrices; therefore the angular part should be precisely that of the Pauli two-component theory. Defining $\chi_+ = \begin{bmatrix} 1 \\ 0 \end{bmatrix}$ and $\chi_- = \begin{bmatrix} 0 \\ 1 \end{bmatrix}$ the spin dependent wavefunctions can be written as [17,18]

$$|n,l,m,\uparrow\rangle \equiv \Psi_{n,j=l+\frac{1}{2},m_j} = \sqrt{\frac{l+m_j+\frac{1}{2}}{2l+1}}R_{nl}(r)Y_l^{m_j-1/2}\chi_+ + \sqrt{\frac{l-m_j+\frac{1}{2}}{2l+1}}R_{nl}(r)Y_l^{m_j+1/2}\chi_-$$

$$\equiv F_1 Y_l^{m_j-1/2}\chi_+ + F_2 Y_l^{m_j+1/2}\chi_- \tag{11}$$

$$|n,l,m,\downarrow\rangle \equiv \Psi_{n,j=l-\frac{1}{2},m_j} = -\sqrt{\frac{l-m_j+\frac{1}{2}}{2l+1}}R_{nl}(r)Y_l^{m_j-1/2}\chi_+ + \sqrt{\frac{l+m_j+\frac{1}{2}}{2l+1}}R_{nl}(r)Y_l^{m_j+1/2}\chi_-$$

$$\equiv -F_2 Y_l^{m_j-1/2}\chi_+ + F_1 Y_l^{m_j+1/2}\chi_-. \tag{12}$$

**4. Magnetic Flux Quantization Argument**

As was stated earlier for calculating the quantized magnetic flux [5], the area integral of the magnetic field must be converted to a time integral over the cyclotron period $T_c$ ($=\frac{2\pi}{\omega_c}=\frac{2\pi m_e}{eB}$). Therefore the important time interval is cyclotron period of the magnetic field :

$$\Phi = \int_S \vec{B}.d\vec{a} = \oint \frac{\vec{B}}{2}.(\vec{r}\times d\vec{r})\frac{dt}{dt} = \oint \frac{\vec{B}}{2}.(\vec{r}\times\dot{\vec{r}})dt = \int_0^{T_c=\frac{2\pi}{\omega_c}} \frac{\vec{B}}{2}.(\vec{r}\times\dot{\vec{r}})dt \tag{13}$$

which will be the starting point for calculating a quantized magnetic flux through any quantum orbit. In the following we calculate the quantized magnetic fluxes through the electron's orbits which are determined by the solutions of the Dirac equation. The source of the magnetic field will be that of proton's magnetic moment $\mu_p$. The probability density $P(r) \approx r^2 |R_{nl}(r)|^2$ has its maxima at $r_n=n^2 a_0$ and in the semiclassical picture electron is assumed to be moving in circular orbits with radius $r_n=n^2 a_0$ around proton which is at rest at the origin of its rest frame. The flux coming from proton's magnetic moment $\mu_p$ is found to depend on the quantum numbers, $n$ and $l$ only : $\Phi_{nl} = (n - l)\Phi_0$. We add to this the fluxes coming from the induced effects of the orbital and spin angular momenta, namely we calculate the similar fluxes due to the orbital and spin magnetic moments ($\mu_l$ and $\mu_s = \mu_e$) of the electron. The last two fluxes are found to depend on the quantum numbers $m_l$ and $m_s$ respectively. So for ($l = n-1$) the resultant quantum flux corresponding to the quantum numbers: $[n,l,m_j]$ is found to be: $\Phi(n,l,m_j) = \Phi_{nl} + \Phi_{m_j} = [n-l-m_j]\Phi_0 = [n-l-m_l-m_s]\Phi_0$. The selection rule [9] for an optical transition ($|n,l,m_j\rangle \rightarrow |n',l',m_j'\rangle$ ; $\Delta m_j = 0, \pm 1$) results that in an optical transition the flux change must be: $\Delta\Phi = 0, \pm \Phi_0$.

**a) Calculation of $\Phi_{nl}$:**

Since the Coulomb potantial requires planar orbits for electron, we assume that proton's magnetic moment $\mu_p$ is in the z-direction and the electron is rotating in (x-y) plane. If the proton's radius is $r_p$, then outside the proton the vector potential is:

$$\vec{A}_{out} = \frac{\vec{\mu}_p \times \vec{r}}{r^3} \qquad (r \geq r_p) \qquad (14)$$

where $\vec{r}$ is the vector going from origin to the point where $\vec{A}$ is being computed. The magnetic field $\vec{B}$ is calculated through the relation:

$$\vec{B}_{out} = \vec{\nabla} \times \vec{A}_{out} \qquad (15)$$

In the x-y plane the x and y components of $\vec{B}_{out}$ vanish and the z component takes the form

$$+(\vec{B}_{out})_z = -\frac{\mu_p}{r^3} \qquad (r \geq r_p) \qquad (16)$$

where $r = \sqrt{x^2 + y^2}$.

First we want to calculate the fluxes through the x-y plane outside and inside the proton: The related fluxes will be defined as follows:

$\Phi_{out}(r_p)$ = The total flux through x-y plane outside the proton ($r \geq r_p$)

$\Phi_{in}(r_p)$ = The total flux through x-y plane inside the proton ($r \leq r_p$)

The condition of no magnetic monopoles $(\vec{\nabla}.\vec{B}=0)$ requires the number of magnetic field lines inside the proton must be equal to the number of magnetic field lines outside the the proton . Therefore

$$\Phi_{in}(r_p) = -\Phi_{out}(r_p). \qquad (17)$$

$\Phi_{out}(r_p)$ can be calculated easily from Eq.(16):

$$\Phi_{out}(r_p) = \int_{r_p}^{\infty} \vec{B}_{out}.d\vec{a} = -\int_{r_p}^{\infty} \frac{\mu_p}{r^3} 2\pi r dr = -\frac{2\pi\mu_p}{r_p} \qquad (18)$$

From Eq.(17) $\Phi_{in}(r_p)$ will be

$$\Phi_{in}(r_p) = \frac{2\pi\mu_p}{r_p}. \qquad (19)$$

Secondly we define $\Phi(r_n)$ which is the flux through the electron orbit with radius $r_n=n^2 a_0$. Since we assume that electron is rotating in x-y plane, $\Phi(r_n)$ will be :

$$\Phi(r_n) = \int_0^{r_n} \vec{B}.d\vec{a} = \int_0^{r_p} \vec{B}_{in}.d\vec{a} + \int_{r_p}^{r_n} \vec{B}_{out}.d\vec{a} \qquad (20)$$

The first term in Eq.(20) is simply $\Phi_{in}(r_n)$, therefore subtitution of Eqs.(19) and (16) in Eq.(20) gives

$$\Phi(r_n) = \frac{2\pi\mu_p}{r_p} - \int_{r_p}^{r_n} \frac{\mu_p}{r^3} 2\pi r dr = \frac{2\pi\mu_p}{r_n} = \frac{2\pi\mu_p}{n^2 a_0} \qquad (21)$$

where we put $r_n=n^2 a_0$.

The result that we found in Eq.(21) is the flux that correspond to one circling period, $T$ .But we know that [5] , for calculating the quantized magnetic flux ,the area integral of the magnetic field must be converted to a time integral over the cyclotron period $T_c$ ($=\frac{2\pi}{\omega_c} = \frac{2\pi m_e}{eB}$). Therefore to find the quantized magnetic flux corresponding to the period $T_c$ we must multiply the flux $\Phi(r_n)$ in Eq.(21) by $\frac{T_c}{T}$ which requires the calculations of $T$ and $T_c$ (or $\omega$ and $\omega_c$) in terms of $r_n=n^2 a_0$. We know that in the hydrogen atom the Coulomb force ($e^2/r^2$) causes electron to rotate around proton. In addition to this we will have the centrifugal force coming from the last term of the Eq.(5). So the net force will be ($\frac{e^2}{r^2} - \frac{l(l+1)\hbar^2}{m_e r^3}$) which results in a zero tork with respect to the origin. Therefore the total angular momentum is a constant of motion: In the semiclassical picture first we set: $l(l+1)\hbar^2 = l^2\hbar^2$ and we denote the inertial moment of the electron with respect to the origin by $I=mr^2$. Then the angular momentum coming from the first term is $L_1=I\omega_1$ with ($\omega_1=\sqrt{\frac{e^2}{m_e r^3}}$) and the angular momentum coming from the second term is $L_2=\hbar l=I\omega_2 \equiv I\omega_l$ with ($\omega_2 \equiv \omega_l = \frac{l\hbar}{m_e r^2}$). Therefore the net angular velocity $\omega$ in terms of $r_n = n^2 a_0$ is: $\omega = \omega_1 - \omega_2$

$$\omega = \omega_1 - \omega_2 = \left[\frac{e^2}{m_e r^3}\right]^{1/2} - \frac{l\hbar}{m_e r^2} = \left[\frac{e^2}{m_e (n^2 a_0)^3}\right]^{1/2} - \frac{l\hbar}{m_e (n^2 a_0)^2} \qquad (22)$$

On the other hand the magnetic field, $B(r_n)$ of the proton at the position of the electron defines another period which is the cyclotron period, $T_c = \frac{2\pi m_e}{eB(r_n)}$. From Eq.(16) $\omega_c$ takes the form:

$$\omega_c = \frac{e\mu_p/r_n^3}{m_e} = \frac{e\mu_p}{m_e (n^2 a_0)^3} \qquad (23)$$

If we calculate $T_c$ and $T$ for $n=1$ we can see that $T_c \approx 10^6 T$. Therefore during the cyclotron period $T_c$ electron rotates $\frac{T_c}{T} \equiv 10^6$ times around proton. As it was stated in Eq.(13) for calculating the quantized magnetic flux, the important time interval is cyclotron period $T_c$ of the magnetic field. Therefore we must multiply the flux $\Phi(r_n)$ in Eq.(21) by $\frac{T_c}{T}$. So from Eqs.(21),(22) and (23), the flux per cyclotron period (the quantum flux) is

$$\Phi_{nl} = \frac{T_c}{T}\Phi(r_n) = \frac{\omega}{\omega_c}\frac{2\pi\mu_p}{n^2 a_0} = (n-l)\frac{hc}{e} = (n-l)\Phi_0 \qquad (24)$$

Eq.(24) is the flux corresponding to the quantum number ($n$ and $l$). In the following we calculate the magnetic fluxes coming from the other quantum numbers $m_j$, hence $m_l$ and $m_s$.

**b) Calculation of $\Phi_{m_j}$:**

We start with the normalization conditions for wave functions $\Psi_{njm_j}$ given in Eq.(11) and (12):

$$\int_0^\infty \int_0^\pi \left|\Psi_{n,j,m_j}(r,\theta,\phi)\right|^2 r^2 \sin\theta \, dr d\theta = \frac{1}{2\pi} \qquad (25)$$

since $\left|\Psi_{n,j,m_j}(r,\theta,\phi)\right|$ is independent of $\phi$. The $z$-component angular momentum operator $\hat{J}_z = i\hbar\partial/\partial\phi$ admits $\Psi_{n,j,m_j}(r,\theta,\phi)$ as an eigenfunction corresponding to the eigenvalue $J_z = m_j\hbar$. As was stated earlier the probability density corresponding to the eigenfunction $\Psi_{n,j,m_j}(r,\theta,\phi)$ will be maximum for $r_n = n^2 a_0$ with $\sin\theta = 1$ in x-y plane. Therefore in the semiclassical picture the electron will execute a circular motion about z-axis. This is born out by the fact that due to the radial wave function and the $\theta$-dependent part of the spherical harmonic being real-valued the probability current density arising from an energy eigenfunction $\Psi_{n,j,m_j}(r,\theta,\phi)$ is zero along the radial and the $\theta$ directions. There is a non-zero probability current density for the circular motion along the $\phi$ direction given by [19,20]

$$j(r,\theta,\phi) = -\frac{i\hbar}{2m_e}\left(\Psi^*_{n,j,m_j}\frac{\partial \Psi_{n,j,m_j}}{a\partial\phi} - \frac{\partial \Psi^*_{n,j,m_j}}{a\partial\phi}\Psi_{n,j,m_j}\right) \quad (26)$$

$$= \frac{J_z}{m_e a}\left|\Psi_{n,j,m_j}(r,\theta,\phi)\right|^2 \quad (27)$$

where $m_e$ is the mass of the electron. The probability current density across an elementary surface area, $rdrd\theta$ perpendicular to the flow is [21]:

$$j(r,\theta,\phi)rdrd\theta = \frac{J_z}{m_e a}\left|\Psi_{n,j,m_j}(r,\theta,\phi)\right|^2 rdrd\theta. \quad (28)$$

A probability current density $j(r,\theta,\phi)$ gives rise to an electric current density [22]

$$j_e(r,\theta,\phi) = -e\,j(r,\theta,\phi). \quad (29)$$

Here the negative sign shows that the electric current flows in an opposite direction to that of the probability current due to the negative nature of the electron charge. It follows that there is an electric current circling the z-axis in a circle of radius $a = r\sin\theta$ flowing perpendicularly across an elementary surface area $rdrd\theta$ given by

$$j_e(r,\theta,\phi)rdrd\theta = -\frac{eJ_z}{m_e a}\left|\Psi_{n,j,m_j}(r,\theta,\phi)\right|^2 rdrd\theta \quad (30)$$

Our next objective is to establish a quantized magnetic flux coming from the current of the orbiting electron (including spin). The idea is based on the familiar notion that an electric current flowing round a circle of radius $a$ about the z-axis would enclose a magnetic flux generated by the current. This flux should be proportional to the current, the proportionality constant being the self-inductance [23,26]. So, a current element in Eq. (30) circulating about the $z$-axis in a circle of radius $a$ should enclose an induced magnetic flux element $d\Phi_z(r,\theta,\phi)$ proportional to the current $j_e(r,\theta,\phi)rdrd\theta$, i.e., we have

$$d\Phi_z = L_e(a) j_e(r,\theta,\phi) rdrd\theta \tag{31}$$

where $L_e(a)$ is the self-inductance. For a quantum current flowing round a circle of radius $a$ with a self-inductance is [26,27]:

$$L_e(a) = m_e \left(\frac{2\pi a}{e}\right)^2. \tag{32}$$

Substitution of Eq.(32) into (31) gives

$$\begin{aligned}d\Phi_z(r,\theta,\phi) &= L_e(a) j_e(r,\theta,\phi) rdrd\theta \\ &= m_e \left(\frac{2\pi a}{e}\right)^2 \left(-\frac{eJ_z}{m_e a}\right) \left|\Psi_{n,j,m_j}(r,\theta,\phi)\right|^2 rdrd\theta \\ &= -\frac{(2\pi)^2 J_z}{e} \left|\Psi_{n,j,m_j}(r,\theta,\phi)\right|^2 r^2 \sin\theta\, drd\theta\end{aligned} \tag{33}$$

the radius $a$ having been replaced by $r\sin\theta$ in the last step. So the total induced quantized magnetic flux of an electron in an energy eigenstate $\Psi_{nlm_j}(r,\theta,\phi)$ will be

$$\begin{aligned}\Phi_z(induced) &= \int_0^\infty \int_0^\pi d\Phi_z^{(0)}(r,\theta,\phi) \\ &= -\frac{(2\pi)^2 J_z}{e} \int_0^\infty \int_0^\pi \left|\Psi_{n,j,m_j}(r,\theta,\phi)\right|^2 r^2 \sin\theta\, drd\theta \\ &= -\frac{2\pi}{e} J_z\end{aligned} \tag{34}$$

Substitution of $J_z = m_j \hbar$ in Eq.(34) we get $\Phi_z(induced) \equiv \Phi_{m_j}$:

$$\Phi_{m_j} = -m_j \Phi_0 \quad ; \quad \Phi_0 = \frac{h}{|e|}. \tag{35}$$

The resulting total quantized magnetic flux through the electron orbits corresponding to ($n,l,m_j$) state has been found to be :

$$\Phi(n,l,m_j) = [n - l - m_j]\Phi_0 = [n - l - m_l - m_s]\Phi_0 . \qquad (36)$$

If we are dealing with dipole transitions, the flux change between two eigenstates is:

$$\Delta\Phi = \Delta n - \Delta l - (\Delta m_l + \Delta m_s)\Phi_0 = (\Delta n - \Delta l - \Delta m_j)\Phi_0 . \qquad (37)$$

As we are dealing with ($l = n$-1) states, we have:

$$\Delta n = \Delta l \qquad (38)$$

So the flux change, $\Delta\Phi$ ( or $\Delta\Phi_z$) becomes :

$$\Delta\Phi \equiv \Delta\Phi_z = -\Delta m_j \Phi_0 . \qquad (39)$$

For a dipole transition the spin dependent selection rules were obtained recently by Saglam et. al. [9] for Dirac hydrogen atom using the well-known Golden rule [10] :

$$\Delta m_j = 0, \pm 1 \qquad (40)$$

Substitution of Eq.(40) in Eq.(39) gives the selection rules for the magnetic flux in a dipole transition:

$$\Delta\Phi = 0, \mp\Phi_0 . \qquad (41)$$

So in the dipole transitions (including optical transitions ) the change in the total angular momentum is associated with a similar change in the quantum flux. In the following section we will apply the present result to the 1s-2p transition of the hydrogen atom.

**5. Aplication to the 1s-2p transition:**

Since we assume that proton's magnetic moment vector $\mu_p$ is in the z-direction, then in the ground state electron will occupy the 1s (spin up) $\equiv |1,0,0,\uparrow>$ state. From Eq.(36), the quantized flux corresponding to the ground state will be:

$$\Phi(1,0,0,1/2)=(1/2)\Phi_0 \qquad (42)$$

We will have 6 different spin dependent 2p states which are: $|2,1,0,\uparrow\downarrow>$ ; $|2,1,1,\uparrow\downarrow>$ and $|2,1,-1,\uparrow\downarrow>$. The corresponding quantized fluxes for these states will be:

$$\Phi(2,1,0,1/2)=(1/2)\Phi_0$$

$$\Phi(2,1,0,-1/2)=(3/2)\Phi_0$$

$$\Phi(2,1,1,1/2)= -(1/2)\Phi_0 \qquad (43)$$

$$\Phi(2,1,1,-1/2)=(1/2)\Phi_0$$

$$\Phi(2,1,-1,1/2)=(3/2)\Phi_0$$

$$\Phi(2,1,-1,-1/2)=(5/2)\Phi_0 \ .$$

Acccording to the selection rule given by Eq.(41) the possible transitions will be as follows:

$$|1,0,0,\uparrow\rangle \quad \to \quad |2,1,0,\uparrow\rangle \ \text{....(allowed, } \Delta\Phi_z=0 \text{ )}$$

$$|1,0,0,\uparrow\rangle \quad \to \quad |2,1,0,\downarrow\rangle \ \text{....(allowed, } \Delta\Phi_z=\Phi_0\text{)}$$

$$|1,0,0,\uparrow\rangle \quad \to \quad |2,1,1,\downarrow\rangle \ \text{....(allowed, } \Delta\Phi_z=0 \text{ )} \qquad (44)$$

$$|1,0,0,\uparrow\rangle \quad \to \quad |2,1,-1,\uparrow\rangle \ \text{....(allowed, } \Delta\Phi_z=\Phi_0\text{)}$$

But the following transitions are not allowed, because, the excited state can not have higher than the ground state flux [5] and the flux change can not be higher than $\Phi_0$.

$$|1,0,0,\uparrow\rangle \quad \overset{X}{\longrightarrow} \quad |2,1,1,\uparrow\rangle ....\text{(not allowed, } \Delta\Phi_z = -\Phi_0 \text{)}$$

$$|1,0,0,\uparrow\rangle \quad \overset{X}{\longrightarrow} \quad |2,1,-1,\downarrow\rangle ....\text{(not allowed, } \Delta\Phi_z = -2\Phi_0\text{)} \tag{45}$$

As it is seen from Eq.(44) some transitions permit spin flip which is in agreement with our earlier result [9] where the related matrix elements were:

$$\langle 2,1,0,\uparrow | -d.E | 1,0,0,\uparrow \rangle = \frac{\sqrt{2}}{3}C$$

$$\langle 2,1,0,\downarrow | -d.E | 1,0,0,\uparrow \rangle = \frac{2}{3}C$$

$$\langle 2,1,1,\downarrow | -d.E | 1,0,0,\uparrow \rangle = -\frac{1}{3}C$$

$$\langle 2,1,-1,\uparrow | -d.E | 1,0,0,\uparrow \rangle = -\frac{\sqrt{2}}{3}C \tag{46}$$

$$\langle 2,1,1,\uparrow | -d.E | 1,0,0,\uparrow \rangle = 0$$

$$\langle 2,1,-1,\downarrow | -d.E | 1,0,0,\uparrow \rangle = 0$$

In the last two lines we have zero matrix elements which means no transitions at all.

## 6. Conclusions

We have calculated the quantized magnetic fluxes through the electronic orbits of Dirac hydrogen atom in the absence of an external magnetic field. The sources of the magnetic field are taken to be that of proton's magnetic moment $\mu_p$ and electron's magnetic moment $\mu_e$ (or $\mu_j$) which has two components namely the orbital part $\mu_l$ and the spinning part $\mu_s$. We show that the quantized magnetic fluxes through the electronic orbits corresponding to the $(n,l,m_j)$ eigenstates can be expressed in tems of the related quantum numbers, namely; $\Phi(n,l,m_j) = [n - l - m_j]\Phi_0 = [n - l - m_l - m_s]\Phi_0$ where $\Phi_0 = \frac{h}{|e|}$ is the flux quanta. We apply the present result to the dipole transitions of hydrogen atom and show that the selection rule ($\Delta m_j = 0, \pm 1$) that were obtained recently by Saglam et. al. [9] can be replaced by $\Delta \Phi = 0, \mp \Phi_0$. So the change in the total angular momentum is associated with a similar change in the quantum flux as well. Since during an optical transition a photon is absorbed or emitted then the above selection rule imply that photon should carry a quantized magnetic flux of $\Phi_0$ with itself. It is believed that the present result will bring a new insight to understant the polarization spectroscopy of hiydrogen-like atoms such as Cs [28]. The application of the present result to the selection rules for the optical transitions of hydrogen atom gives access to the spin flip-floppings. Therefore it is also believed to serve a significant help for understanding of the recent observations of spin relaxation in excitonic transitions (such as 1s→2p or 2p→3d ) in nanostructures [11-13].